\documentclass[prl,amssymb,amsmath,aps,11pt,tightenlines,superscriptaddress,longbibliography,notitlepage]{revtex4-1}

\usepackage{graphicx,mdwlist,color,wrapfig,revsymb}
\usepackage[us]{datetime}

\begin{document}

\title{Induced superconductivity in high mobility two dimensional electron gas in GaAs heterostructures.}

\author{Zhong Wan}
\affiliation{Department of Physics and Astronomy, Purdue University, West Lafayette, IN 47907 USA}
\author{Aleksandr Kazakov}
\affiliation{Department of Physics and Astronomy, Purdue University, West Lafayette, IN 47907 USA}
\author{Michael J. Manfra}
\affiliation{Department of Physics and Astronomy, Purdue University, West Lafayette, IN 47907 USA}
\affiliation{Department of Electrical Engineering, Purdue University, West Lafayette, IN 47907 USA}
\affiliation{Department of Materials Engineering, Purdue University, West Lafayette, IN 47907 USA}
\affiliation{Birck Nanotechnology Center, Purdue University, West Lafayette, IN 47907 USA}
\author{Loren N. Pfeiffer}
\affiliation{Department of Electrical Engineering, Princeton University, Princeton, NJ 08544 USA}
\author{Ken W. West}
\affiliation{Department of Electrical Engineering, Princeton University, Princeton, NJ 08544 USA}
\author{Leonid P. Rokhinson}
\email{leonid@purdue.edu}
\affiliation{Department of Physics and Astronomy, Purdue University, West Lafayette, IN 47907 USA}
\affiliation{Department of Electrical Engineering, Purdue University, West Lafayette, IN 47907 USA}
\affiliation{Birck Nanotechnology Center, Purdue University, West Lafayette, IN 47907 USA}


\maketitle


\textbf{Introduction of a Josephson field effect transistor (JoFET) concept \cite{Clark1980} sparked active research on proximity effects in semiconductors. Induced superconductivity and electrostatic control of critical current has been demonstrated in two-dimensional gases in InAs\cite{Takayanagi1985,Akazaki1996}, graphene\cite{Heersche2007} and topological insulators\cite{Sacepe2011,Williams2012,Veldhorst2012,Qu2012,Pribiag2014}, and in one-dimensional systems\cite{Doh2005,Jarillo-Herrero2006,Xiang2006} including quantum spin Hall edges\cite{Hart2013,Yu2014}. Recently, interest in superconductor-semiconductor interfaces was renewed by the search for Majorana fermions\cite{Rokhinson2012a,Mourik2012}, which were predicted to reside at the interface\cite{Fu2009,Lutchyn2010a,Alicea2010}. More exotic non-Abelian excitations, such as parafermions (fractional Majorana fermions)\cite{Clarke2012,Mong2013,Zuo2014} or Fibonacci fermions may be formed when fractional quantum Hall edge states interface with superconductivity. In this paper we develop transparent superconducting contacts to high mobility two-dimensional electron gas (2DEG) in GaAs and demonstrate induced superconductivity across several microns. Supercurrent in a ballistic junction has been observed across 0.6 $\mu$m of 2DEG, a regime previously achieved only in point contacts but essential to the formation of well separated non-Abelian states. High critical fields ($>16$ Tesla) in NbN contacts enables investigation of a long-sought regime of an interplay between superconductivity and strongly correlated states in a 2DEG at high magnetic fields\cite{Zyuzin1994,Fisher1994,Hoppe2000,Kim2004,Giazotto2005,vanOstaay2011}.}

Proximity effects in GaAs quantum wells have been intensively investigated in the past and Andreev reflection has been observed by several groups\cite{Lenssen1993,Moore1999,Verevkin1999,Takayanagi2002}. Unlike in InAs, where Fermi level ($E_F$) at the surface resides in the conduction band, in GaAs $E_F$ is pinned in the middle of the gap which results in a high Schottky barrier between a 2DEG and a superconductor and low transparency non-ohmic contacts. Heavy doping can move $E_F$ into the conduction band and, indeed, superconductivity has been induced in heavily-doped bulk n$^{++}$ GaAs\cite{Kutchinsky2001}. In quantum wells similar results were obtained by annealing indium contacts\cite{Marsh1994}, however the critical field of indium is $\sim30$ mT which is well below the fields where quantum Hall effect is observed.

\begin{figure}[t]
\includegraphics[width=0.6\textwidth]{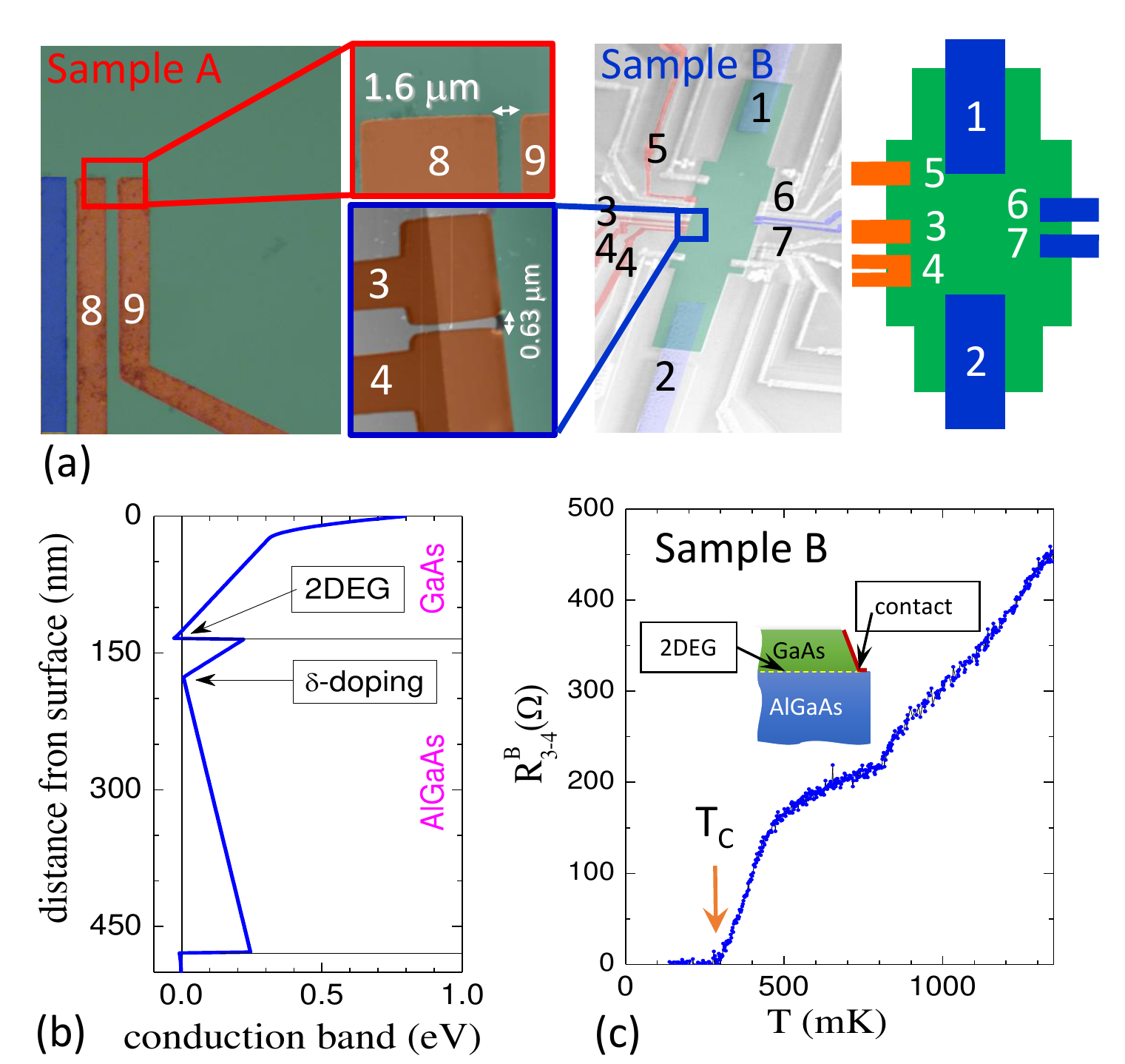}
\caption{{\bf Devices design and superconducting transition.} (a) Scanning electron microscope (SEM) images of test devices similar to samples A and B. Enlarged region for sample B is an atomic force microscope (AFM) image of a real sample. 2D gas regions are false-color coded with green, superconducting and normal contacts are coded with orange and blue, respectively. (b) Simulation of the conduction band energy profile in the heterostructure\cite{snidernote,snider90}. (c) $T$-dependence of resistance between contact 3 and 4 in Sample B measured with 10 nA ac excitation. Superconducting transition is observed at $T_c\approx 290$ mK.}
\label{f-devices}
\end{figure}

In conventional quantum well structures AlGaAs barrier between 2D electron gas (2DEG) and the surface of the sample adds an extra 0.3 eV to the Schottky barrier when contacts are defused from the top. We alleviated these problems by growing an inverted heterojunction structures, where a 2DEG resides at the GaAs/AlGaAs interface but the AlGaAs barrier with modulation doping is placed below the 2DEG, see Fig.~\ref{f-devices}. Contacts are recessed into the top GaAs layer in order to bring superconductor closer to the 2DEG. A thin layer of AuGe and NbN superconductor form low resistance ohmic contacts to the 2DEG after annealing. The inverted heterostructure increases the contact area of side contacts compared to quantum well structures by utilizing all GaAs layer above the heterointerface for carrier injection ($130$ nm in our inverted heterostructure vs $20-30$ nm in typical quantum wells).

Here we report induced superconductivity in two devices from different wafers, sample A has long ($70$ $\mu$m) contacts separated by $1.6$ $\mu$m of 2DEG, for sample B contacts are formed to the edge of a mesa with $0.6$ $\mu$m separation.
Details of device fabrication are described in Methods. When cooled down to 4 K in the dark both samples show resistance in excess of $1$ M$\Omega$. After illumination with red light emitting diode (LED) a 2DEG is formed and 2-terminal resistance drops to $<500\Omega$. As shown in Fig.~\ref{f-devices}d sample resistance $R^B_{3-4}$ gradually decreases upon cooldown from 4 K to the base temperature and the S-2DEG-S junctions becomes superconducting at $T_c\sim 0.3$ K.

\begin{figure}[t]
\includegraphics[width=0.6\textwidth]{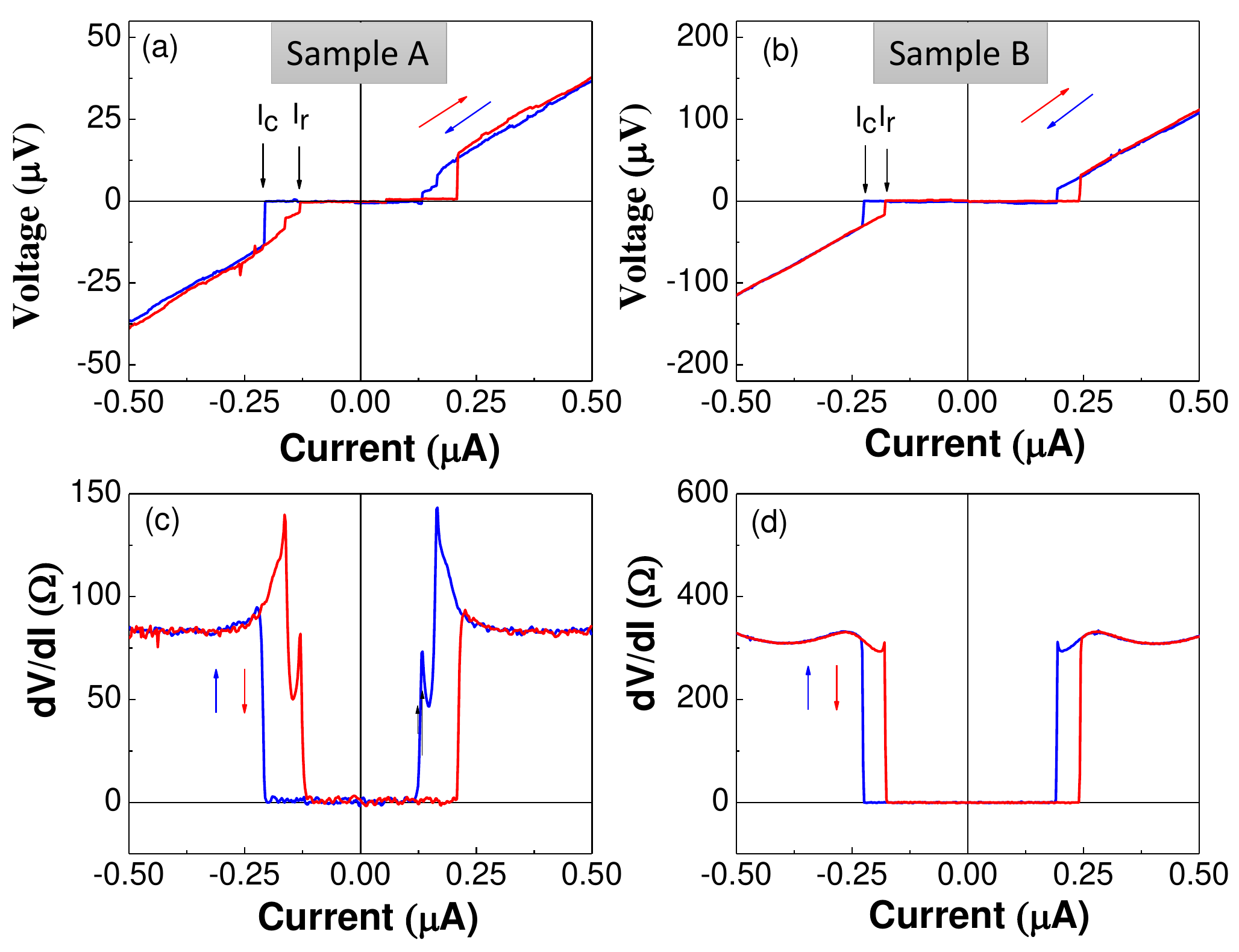}
\caption{{\bf Induced superconductivity in a high mobility 2D electron gas in GaAs.} Voltage-current characteristics and differential resistance are measured between 8-9 for sample A and between 3-4 for sample B at base temperature , $dV/dI$ is measured with $I_{ac}=1$ nA. Induced superconductivity with zero voltage is observed with critical currents $I_c\sim 220$ nA for sample A and $I_c\sim 230$ for sample B.}
\label{f-IV}
\end{figure}

Voltage-current $V(I)$ characteristics for two S-2DEG-S junctions (between 8-9 for sample A and between 3-4 for sample B) are shown in Fig.~\ref{f-IV}. Both samples show zero resistance state at small currents with abrupt switching into resistive state at critical currents $I_c=0.22\ \mu$A and $0.23\ \mu$A for samples A and B respectively. $V(I)$ characteristics are hysteretic most likely due to the Joule heating in the normal state.

\begin{figure}[t]
\includegraphics[width=0.6\textwidth]{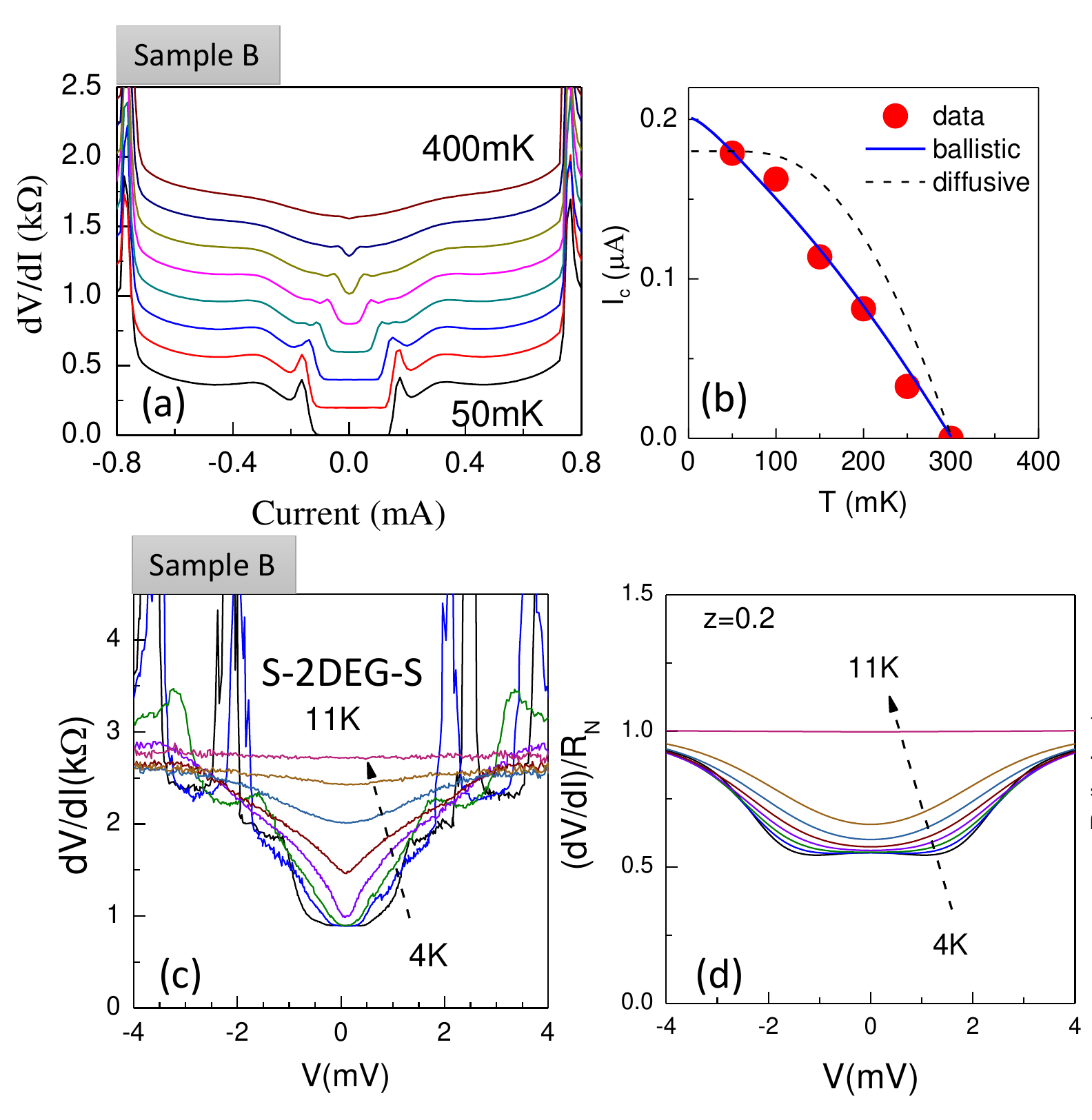}
\caption{{\bf Temperature dependence of superconductivity in a ballistic junction.} (a) Evolution of the induced superconductivity with $T$ for sample B. The $R(I)$ curves are offset proportional to $T$ for $T>50$ mK. (b) Temperature dependence of critical current $I_c(T)$ is extracted from (a) and compared to the expected $T$-dependence for ballistic and diffusive regimes (reduced $I_c$ compared to Fig.~\ref{f-IV} is due to larger $I_{ac}=10$ nA used in this experiment). (c) High-temperature data shows Andreev reflection (excess current and reduced $dV/dI$ around $V=0$. The curves are not offset. In (d) excess current is modeled within the BTK theory\cite{Blonder1982} with $Z=0.2$.}
\label{f-Tdep}
\end{figure}

The most attractive property of a high mobility 2DEG is large mean free path $l\gg\xi_0$, with $l=24\ \mu$m and the BCS coherence length $\xi_0=\hbar v_f/\pi\Delta=0.72\ \mu$m for sample B. Here $v_f=\hbar\sqrt{2\pi n}/m$ is the Fermi velocity, $n$ is a 2D gas density, $m$ is an effective mass, and $\Delta=1.76 k_B Tc=46\ \mu$eV is the induced superconducting gap. Evolution of $V(I)$ with $T$ is shown Fig.~\ref{f-Tdep}a. Experimentally obtained $T$-dependence of $I_c$ is best described by the Kulik-Omelyanchuk theory for ballistic junctions ($L\ll l$)\cite{Kulik1977}, the blue curve on Fig.~\ref{f-Tdep}b. For comparison we also plot $I_c(T)$ dependence for the dirty limit $L\ll\sqrt{l\xi_0}$ \cite{Kulik1975}, which exhibits characteristic saturation of $I_c$ at low temperatures.

In short ballistic junctions $L\ll \xi_0\ll l$  the product $I_c(0) R_N=\pi\Delta/e$ does not depend on the junction length $L$. For $L\sim\xi_0$ this product is reduced by a factor $2 \xi_0/(L+2\xi_0)$  \cite{Bagwell1992}. The measured $I_c R_N = 83\ \mu$V for sample B is in a good agreement with an estimate $\pi\Delta/e\cdot2 \xi_0/(L+2\xi_0)=90\ \mu$V. For sample A the $I_c R_N = 19\ \mu$V while the estimated product is $\approx 50\ \mu$V. The reduction is consistent with the geometry of sample A, where a region of the 2DEG with induced superconductivity is shunted by a large region of a 2DEG in a normal state.

\begin{figure}[t]
\includegraphics[width=0.6\textwidth]{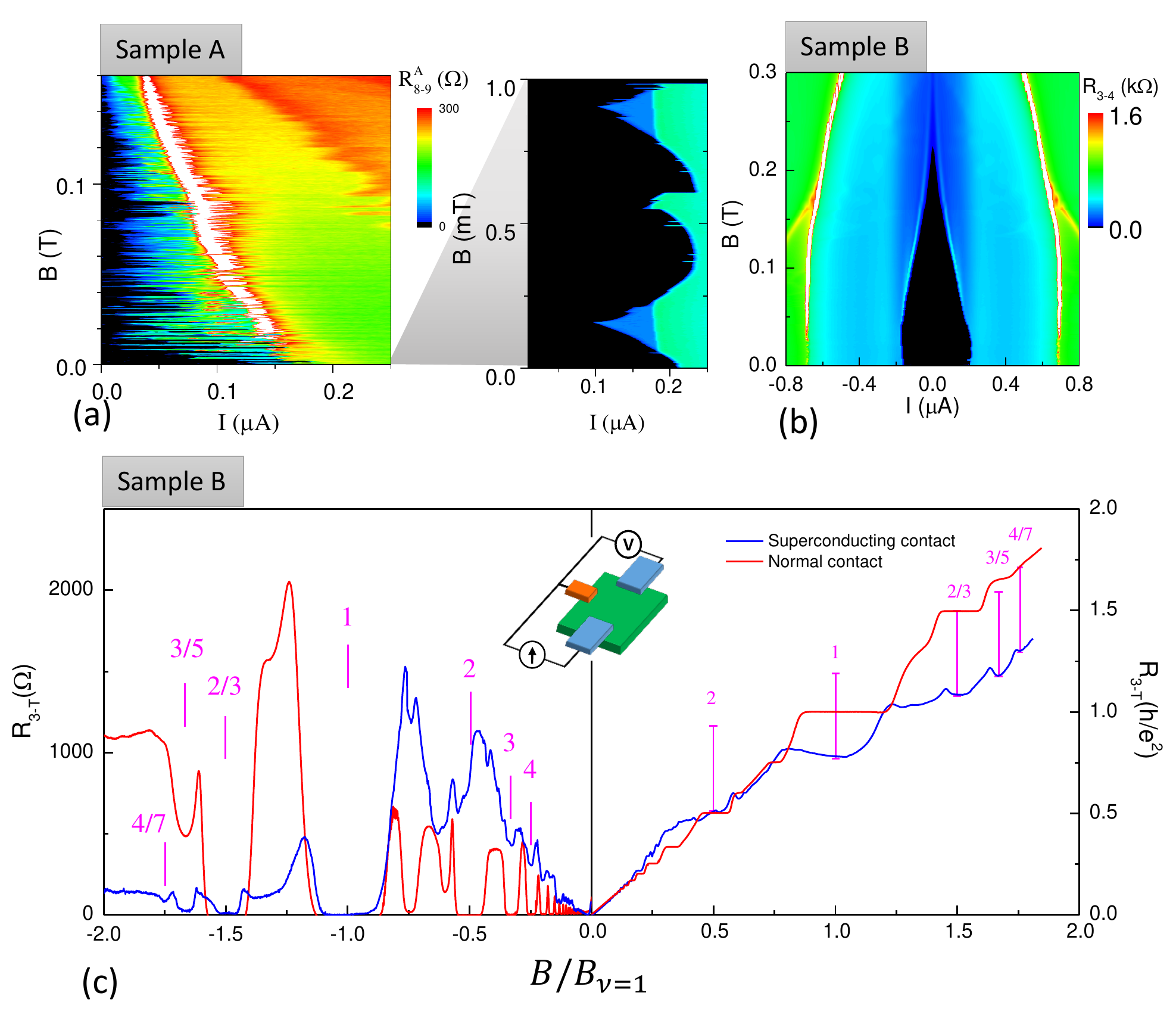}
\caption{{\bf Magnetic field dependence of induced superconductivity.} (a,b) Differential resistance is measured as a function of $B$ and $I_{dc}$ for two samples at 40 mK. Induced superconductivity (black region) is observed up to 0.2 Tesla in both sample. (c) 3-terminal resistance for a sample with all normal contacts (red) and between normal and superconducting contacts in sample B [$I$ $(2-4)$ and $V$ $(4-1)$ in Fig.~\ref{f-devices}] is measured at 70 mK and 40 mK respectively. $B<0$ ($B>0)$ induces clockwise (counterclockwise) chiral edge channels, note resistance scales difference for two field directions.}
\label{f-Bdep}
\end{figure}

Transparency of superconducting contacts can be estimated from the suppression of the superconducting gap in the S-2DEG-S junction between 3-4 in sample B. In one-dimensional junctions the induced gap $\Delta=\Delta_{0}\frac{\Gamma}{\Gamma+\Delta_{0}}$ depends on the broadening of Andreev levels within the semiconductor\cite{Sau2012} $\Gamma=\frac{\hbar v_f}{L_{eff}} D_1D_2$, where we introduce contacts transparencies $D_1$ and $D_2$. We assume for simplicity that $D_1=D_2=1/(1+Z^2)$, where $0<Z<\infty$ is a interface barrier strength introduced in \cite{Blonder1982}, and Bagwell's effective channel length $L_{eff}=L+2\xi_0$. Using NbN superconducting gap $\Delta_{0}=2.02 k_B T_C^{0}$ (NbN is a strong-coupling superconductor, $T_C^{0}=11$ K) we obtain $Z=0.2$. This value is consistent with the fit of the $I_c$ vs $T$ dependence with $D$ as a free parameter, see Supplementary Material for details. Similar values of $Z$ can be estimated from the analysis of the shape of $dI/dV(V)$ characteristics at elevated temperatures, as shown in Fig.~\ref{f-Tdep}. At $T<T_c^0$ Andreev reflection at S-2DEG interfaces results in an excess current flowing through the junction for voltage biases within the superconducting gap $\Delta_0/e$ and corresponding reduction of a differential resistance $dV/dI$ by a factor of 2. In the presence of a tunneling barrier normal reflection competes with Andreev reflection and reduced excess current near zero bias, resulting in a peak in differential resistance. Within the BTK theory\cite{Blonder1982} a flat $dV/dI(V)$ within $\Delta_0/e$, observed in our experiments, is expected only for contacts with very high transparency $Z<0.2$. For larger $Z>0.2$ a peak at low biases is expected (see Supplementary Material). Several features of the experimental $I(V)$ need to be mentioned. First, we observe several sharp peaks in the resistance at high biases (around 2 mV and 4 mV for $T=4$ K). Similar sharp resonances has been observed previously \cite{gao1994}, where authors attributed their appearance to the formation of Fabry-P\'erot resonances between superconducting contacts. In our devices the superconducting region is shunted by a low resistance ($<100 \Omega$) 2DEG, thus appearance of $>10$ k$\Omega$ resonances cannot be explained by resonant electron trapping between contacts. These resonances are also observed in $I(V)$ characteristics of a single S-2DEG interface (measured in the S-2DEG-N configuration between contacts 3-6, see Supplementary Material Fig.~\ref{f-Zfit}). Differential resistance does not change substantially across resonances, ruling out transport through a localized state. We speculate that in the contacts where these resonances are observed superconductivity is carried out by quasi-1D channels, and jumps in I/V characteristics are due to flux trapping at high currents. This scenario is consistent with the observation that peaks shift to lower currents at higher fields, see Fig.~\ref{f-Bdep}. The second notable feature of our data is reduction of the zero-bias resistance by $\approx 2.6$ at low temperatures, while Andreev reflection limits the reduction to the factor of 2. We attribute this reduction to the multiple Andreev reflection between two closely-spaced contacts, for contacts with larger separation (~$20\ \mu$m) multiple Andreev reflection is suppressed and the reduction of resistance by a factor of 2 is observed, see Supplementary Material Fig.~\ref{f-Zfit}.

Finally, we present magnetic field dependence of induced superconductivity. The low-field data is shown in Fig.~\ref{f-Bdep}(a,b), where black regions correspond to zero differential resistance. Induced superconductivity is suppressed at $\approx 0.2$ T in both samples. In sample A a narrow region of a 2DEG with induced superconductivity is confined between large NbN superconducting leads with rigid phases. Perpendicular magnetic field twists the phase in the 2DEG resulting in Fraunhofer-like oscillations of the critical current. In this sample, though, the 2DEG extends beyond the narrow region between the contacts and $I_c$ does not decrease to zero and abrupt jumps in $I_c$ reflect multiple flux jumps. The period of oscillations is $\sim 0.5$ mT which corresponds to an area of 4.1 $\mu$m$^2$, much smaller than the area of the 2DEG between the contacts ($\approx 120$ $\mu$m$^2$). This observation is consistent with the reduced $I_cR_N$ product measured for this sample as discussed above. In sample B contacts are fabricated along the edge of the mesa and 2D gas is not enclosed between the contacts. Consequently, $I_c$ is a smooth function of $B$.

Competition between superconductivity and chiral quantum Hall edge states is shown in Fig.~\ref{f-Bdep}c, where resistance is measured in a 3-terminal configuration over a wide range of magnetic fields. Simple Landauer-Buttiker model of edge states predicts zero resistance for negative and quantized Hall resistance for positive field direction for IQHE and FQHE states, which is clearly seen in a sample with all normal ohmic contacts (red curve). When a superconducting contact serves as a current injector (blue curve), integer $\nu=1$ and fractional $\nu=2/3$ and 3/5 states are well developed for $B<0$, while the same states are not quantized at proper QHE values for $B>0$. If we assume that current injection via a superconducting contact results in an extra voltage offset at the contact $V_{off}\approx\Delta_{ind}/e$, the measured voltage will be reduces by $V_{off}$. The magenta bars for $B>0$ indicate corrected resistance $(V-V_{off})/I$ for $V_{off}=140$ $\mu$V. While this offset may explain the measured values for fractional states, a twice smaller $V_{off}$ is needed to reconcile the resistance at $\nu=1$. Note that induced gap is smaller at higher $B$. At low fields states $\nu=3$, 4 and 5 have resistance minima for $B<0$ indicating partial equilibration of chiral edge currents with the superconducting contact, while resistance near $\nu=2$ has a maximum. Zero resistance at $\nu=1$ and large resistance at $\nu=2$ are in contrast to the theoretical prediction that $\nu=2$ state should be stronger coupled to a superconducting contact than $\nu=1$ \cite{Fisher1994}.

\vspace{0.5cm}
\textbf{Methods}
\vspace{0.2cm}

The GaAs/AlGaAs inverted heterojunctions were grown by molecular beam epitaxy (MBE) on semi-insulating (100) GaAs substrates with the heterointerface placed 130 nm below the surface and $\delta$-doping layer 30-40 nm below the GaAs/AlGaAs interface. Samples were fabricated from two wafers with density and mobility $n=2.7\times10^{11}$ cm$^{-2}$, $\mu=2\times10^{6}$ V$\cdot$s/cm$^{2}$ (sample A) and $n=1.7\times10^{11}$ cm$^{-2}$, $\mu=4\times10^{6}$ V$\cdot$s/cm$^{2}$ (sample B). Superconducting contacts were defined by standard electron beam lithography. First, a 120 nm - deep trench was created by wet etching. Next, samples were dipped into HCl:H$_{2}$O $(1:6)$ solution for 2 s and loaded into a thermal evaporation chamber, where Ti/AuGe (5nm/50nm) was deposited. Finally, 70 nm of NbN was deposited by DC magnetron sputtering in Ar/N$_2$ ($85\%/15\%$) plasma at a total pressure of 2 mTorr. The deposition conditions were optimized for producing high quality NbN films ($T_c=11$ K and $B_c>15$ Tesla) with minimal strain\cite{Glowacka2014}. The contacts were annealed at $500^{\circ}$ C for 10 min in a forming gas. The measurements were performed in a dilution refrigerator with base temperature $<30$ mK, high temperature data was obtained in a variable temperature $^3$He system. Samples were illuminated with red LED at 4 K in order to form a 2D gas, 2-terminal resistance drops from $>1$M$\Omega$ before illumination to $<500$ $\Omega$ after illumination.

\vspace{0.5cm}
\textbf{Acknowledgements}\\
The work at Purdue was supported by the National Science Foundation grant DMR-1307247 (Z.W. and L.P.R.), by the Purdue Center for Topological Materials (Z.W.), and by the U.S. Department of Energy, Office of Basic Energy Sciences, Division of Materials Sciences and Engineering under Awards DE-SC0008630 (A.K.) and  DE-SC0006671 (M.J.M.). The work at Princeton was funded by the Gordon and Betty Moore Foundation through Grant GBMF 4420, and by the National Science Foundation MRSEC at the Princeton Center for Complex Materials.

\vspace{0.5cm}
\textbf{Authors contribution}\\
L.P.R. and M.J.M conceived the experiments, Z.W. fabricated samples, Z.W. and L.P.R performed experiments, Z.W. and L.P.R wrote the manuscript with comments from M.J.M, L.N.P. and K.W.W. designed and grew wafers, A.K. contributed to the fabrication and low temperature experiments.

\vspace{0.5cm}
\textbf{Additional information}\\
Supplementary information is available in the online version of the paper. Correspondence
should be addressed to L.P.R.

\vspace{0.5cm}
\textbf{Competing financial interests}\\
The authors declare no competing financial interests.

\clearpage
\bibliography{rohi}



\renewcommand{\thefigure}{S\arabic{figure}}
\renewcommand{\thetable}{S\arabic{table}}
\renewcommand{\theequation}{S\arabic{equation}}
\renewcommand{\thepage}{sup-\arabic{page}}
\setcounter{page}{1}
\setcounter{equation}{0}
\setcounter{figure}{0}

\begin{center}
\textbf{\Large Supplementary Materials} \\
\vspace{0.2in} \textsc{Superconductivity in ballistic junctions in high mobility two dimensional electron gas in GaAs heterostructures.}\\
{\it Zhong Wan, Michael Manfra, Loren Pfeiffer, Ken West, and Leonid P. Rokhinson}
\end{center}

\section{Temperature dependence of the critical current}
\label{transparancy}

\begin{figure}[h]
\includegraphics[width=0.9\textwidth]{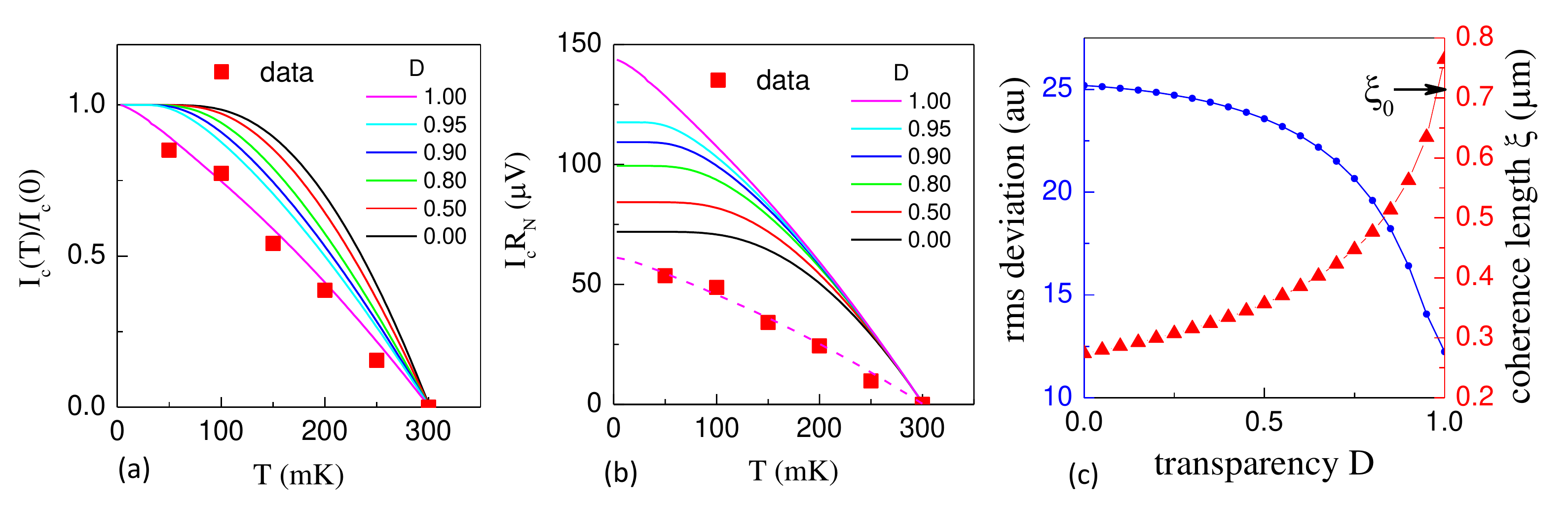}
\caption{{\bf Analysis of the temperature dependence of the critical current.} Scaled (a) and unscaled (b) product $I_cR_N$ is calculated using Eq. (\ref{eq:IcwithD}) for different transparencies $D$ and $\alpha=1$. Red dots are experimental data. Dashed line in (b) is for $\alpha=0.7$ and $D=1$. In (c) root-mean-square deviation between the best fit and the experimental data is shown for different $D$, coherence length $\xi$ obtained from the best fit are red triangles.}
\label{f-Dfit}
\end{figure}

Haberkorn et al. \cite{Haberkorn1978} generalized  Kulik-Omelyanchuk current-phase relations\cite{Kulik1977,Kulik1975} to the case of arbitrary transparency of a tunnel barrier $D$ inserted into the Josephson junction by directly solving Gor'kov's equations. They obtain the following current-phase relation:
\begin{equation}
I_s(\phi,T)R_N=\alpha\frac{\pi\Delta(T)}{2e}\frac{\sin(\phi)}{\sqrt{1-D\sin^2(\phi/2)}} \times
\tanh\frac{\Delta(T)}{2k_BT}\sqrt{1-D\sin^2(\phi/2)},
\label{eq:IcwithD}
\end{equation}
where $\Delta(T)$ is the BCS gap. For $\alpha=1$ this equation interpolated between diffusive ($D=0$) and ballistic ($D=1$) junctions. Critical current can be found as $I_c(T)R_N=max[I_s(\phi,T)R_N]$. We introduce coefficient $\alpha$ to account for the reduction of the critical current due to the finite length of the junction $L$, $\alpha=2\xi/(L+2\xi)$ \cite{Bagwell1992}. The best fit of the experimental $I_cR_N(T)$ dependence assuming both $\alpha$ and $D$ as free parameters is obtained for $D=1$ and $\alpha=0.7$, see Fig.~\ref{f-Dfit}(a,b).  For the contact spacing $L=0.63\ \mu$m this $\alpha$ corresponds to $\xi=0.76\ \mu$m, consistent with the BCS coherence length $\xi_0=\hbar v_f/\pi\Delta=0.72\ \mu$m. Transparency $D$ can be related to the dimensionless barrier strength $Z$ introduced in the Blonder-Tinkham-Klapwijk (BTK) theory\cite{Blonder1982}, $D=1/(1+Z^2)$, and the fit sets the upper limit on $Z$, $Z<0.1$. The quality of the fit parameters can be assessed from Fig.~\ref{f-Dfit}(c), where rms error for the best fit with a fixed $D$ and $\alpha$ as a free parameter $(\mathrm{rms\  deviation})^2=\sum_i\{[I_c(T_i)R_N]^{theory}-[I_c(T_i)R_N]^{exp}\}^2$ is plotted for different $D$. The rms deviation has a clear global minimum at $D\rightarrow1$. Note that the coherence length for $D<1$, obtained from the fitting parameter $\alpha$, becomes smaller than the estimated $\xi_0$.

\section{Analysis of excess current above the induced superconductivity gap}
\label{BTK theory}

\begin{figure}[h]
\includegraphics[width=0.9\textwidth]{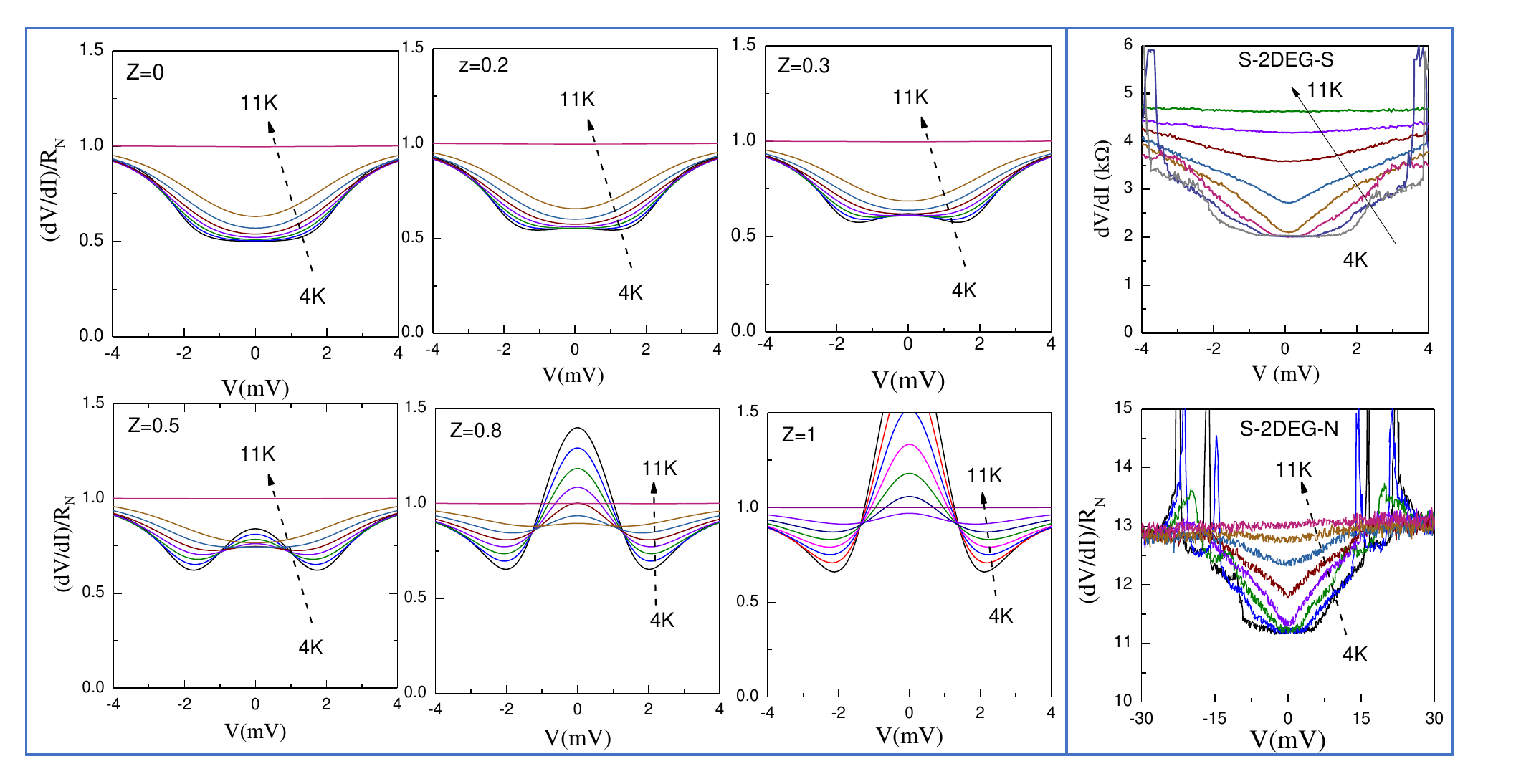}
\caption{{\bf Temperature dependence of differential resistance.} Left 6 plots: normalized differential resistance is calculated using BKT theory, Eq.~\ref{eq:BTK} for different barriers $Z$ and temperatures between 4 and 11 K with a step of 1 K. Right 2 plots: experimentally measured differential resistance between two superconducting contacts ($R_{3-5}$) and a normal-superconducting contact ($R_{4-7}$) in sample B  (the normal contact has high resistance).}
\label{f-Zfit}
\end{figure}

Transparency of the superconductor-semiconductor interface can be estimated from the shape of the $dV/dI(V)$ characteristic, where competition between Andreev and normal reflections results in a peak in differential resistance when a tunneling barrier is present at the superconductor-semiconductor interface (transmission $D=1/(1+Z^2)<1$). Differential resistance for different temperatures can be calculated using Blonder-Tinkham-Klapwijk (BTK) theory\cite{Blonder1982}:
\begin{equation}
\frac{dI}{dV}(V)\propto\int^{\infty}_{-\infty}\frac{\partial{f}_{0}(E-eV)}{\partial(eV)}[1+A(E)-B(E)]dE,
\label{eq:BTK}
\end{equation}
where $f_0(E)$ is the Fermi Dirac function and $A(E)$ and $B(E)$ are energy-dependent Andreev and normal reflection coefficients, respectively. Both coefficients depend on the gap of NbN $\Delta_{0}=\Delta(T)$ with $T_{c}^{0}=11$ K and the interface barrier strength $Z$. In Fig.~\ref{f-Zfit} we plot differential resistance for different values of $Z$. At low $T$ for $Z=0$ the barrier is transparent ($D=1$) and all incident electrons are Andreev reflected, which leads to the a reduction of differential resistance by a factor of 2 within the energy gap $\Delta_0$. When $Z$ is finite, part of the incident electrons undergoes normal reflection which results in the increase of the  resistance within the gap.

The exact shape of experimental curves differ from the shape predicted by the BKT theory, the most important deviation being sharp minima near $V=0$ observed at $T$ close to $T_c^{0}$ as compared to a much smoother BKT dependence. To account for a similar sharpening of a zero-bias peak in less transparent contacts ($Z>2$) it has been assumed that a thin normal region is formed between NbN contacts and a 2DEG\cite{Neurohr1996}. This more elaborate theory introduces two more fitting parameters for the superconducting-normal and normal-2DEG interfaces, but does not change the main qualitative prediction of a simpler BTK theory: appearance of a peak near $V=0$ for $Z>0.2$ in $dV/dI(V)$ characteristics.

Experimentally, we observe no zero-bias peak in $dV/dI(V)$ characteristics measured between two superconducting contacts $R_{3-4}$ (S-2DEG-S) or between superconducting and normal contacts $R_{8-9}$ (S-2DEG-N), see Fig.~\ref{f-Tdep} and \ref{f-Zfit}, thus we can set an upper limit $Z<0.2$ and lower limit $D>0.96$ for our contacts.

\section{Comparison between conventional heterostructures and inverted single interface heterostructure}
\label{transparancy}
\begin{figure}[h]
\includegraphics[width=0.9\textwidth]{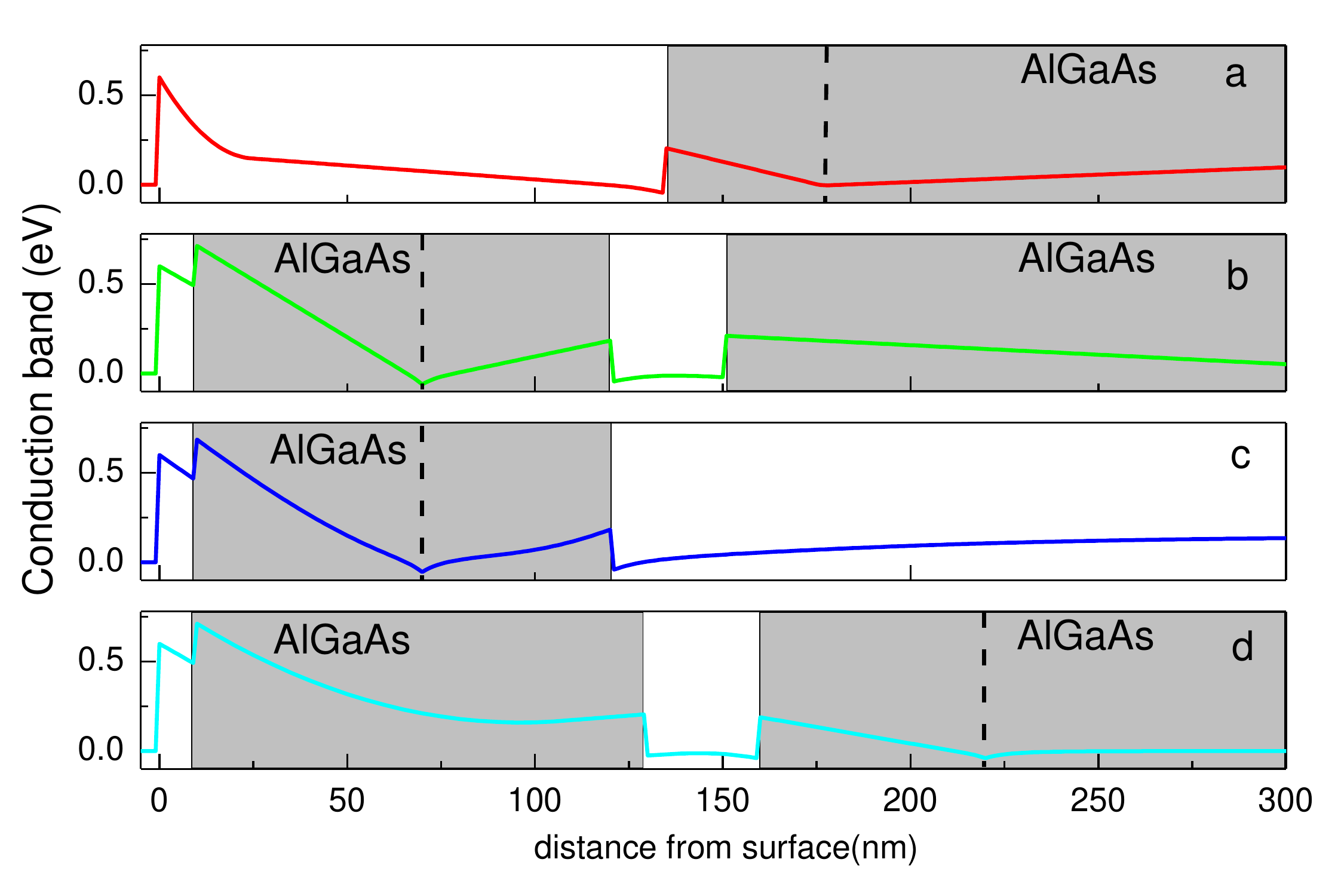}
\caption{{\bf Comparison between conventional heterostructure and inverted single interface heterojunction.} Conduction band profile is plotted for (a) inverted single interface heterojunction used in our experiments and typical (b) modulation-doped quantum well, (c) single heterojunction, and (d) inverted quantum well. Dash lines indicate position of modulation doping.}
\label{f-structure}
\end{figure}

Comparison between conventional heterostructures and inverted heterostructures used in this work is shown in Fig. \ref{f-structure}. In conventional quantum well (b,d) and single interface heterostructures AlGaAs barrier between 2D gas and the surface adds 0.3 eV to the Schottky barrier if contacts are defused from the surface. For side contacts inverted single heterointerface (a) increases the exposed GaAs cross section for Cooper pair injection.

\end{document}